\documentclass[aps,prb,preprint,showpacs,groupedaddress,endfloats]{revtex4}
\usepackage{amssymb}

\usepackage{graphicx}
\usepackage{setspace}
\usepackage{bm}

\newcommand{\be}{\begin{equation}}
\newcommand{\ee}{\end{equation}}
\newcommand{\bd}{\begin{displaymath}}
\newcommand{\ed}{\end{displaymath}}
\newcommand{\ba}{\begin{array}}
\newcommand{\ea}{\end{array}}
\newcommand{\bq}{\begin{eqnarray}}
\newcommand{\eq}{\end{eqnarray}}

\begin{document}

\title{Radiation damage to DNA: electron scattering from the backbone subunits}
\author{Stefano Tonzani}
\affiliation{JILA and Department of Chemistry,  University of Colorado, Boulder, Colorado 80309-0440}
%\ead{tonzani@colorado.edu}
\author{Chris H. Greene}
\affiliation{Department of Physics and JILA, University of Colorado, Boulder, Colorado 80309-0440}

\date{\today}
%\maketitle

\begin{abstract}
In the context of damage to DNA by low-energy electrons, we carry out calculations of electron scattering from tetrahydrofuran and
phosphoric acid, models of the subunits in the DNA backbone, as a first step
in simulating the electron capture process that occurs in the cell. In the case
of tetrahydrofuran, we also compare with previous theoretical and experimental
data. A comparison of the shape of the resonant structures to virtual orbitals
is also performed to gain insight into the systematic connections with electron scattering
from similar molecules and dissociative electron attachment experiments.
\end{abstract}

\maketitle
\section{Introduction}
In the past few years, a growing literature has emerged, concerning the damage to nucleic acids by
low energy electrons
\cite{Sanche:DNA,Sanche_bond:JCP06,Burrow:PRL04,Scheier:AngChem06} produced by
ionizing radiation. The process follows
the creation of thousands of low energy electrons, stripped off
from molecules in the cell either directly by the radiation  or else by its first products, highly
energetic primary electrons, that can cause  electron-impact ionization. It has
been shown that a majority of the secondary electrons
have an energy distribution between 0 and 20 eV. \cite{Cobut:RPC98} If the
electron energy is higher than the ionization threshold for DNA (between 7.85
and 9.4 eV, as measured for the DNA bases \cite{NIST:webbook}), then
the nucleic acid  can be ionized and  damage produced through the subsequent
rearrangement and dissociation of the cation.\cite{rad_damage_DNA:book} If the
electron energy is lower, though, damage can still be generated through a negative
anion-mediated mechanism, which starts with the capture of the electron in a
molecular resonance, followed by a transfer of energy and electron density towards a
weak bond that subsequently ruptures. Beyond this generally accepted framework,
there are many controversial issues.
%, including: Where is the initial capture
%site?
%\cite{Sanche:DNA} How does the metastable
%anion generated by electron capture evolve?  Which is the bond that
%ruptures?\cite{Barrios:JPCB02,Sanche_bond:JCP06} 
These issues concern the location of the initial capture site,\cite{Sanche:DNA} the dynamics of the
metastable anion generated by electron capture and the identification of the
final bond that ruptures.\cite{Barrios:JPCB02,Sanche_bond:JCP06}
The
first question is generally answered saying that since the DNA bases have
extended conjugated systems, they also have many unoccupied $\pi^{*}$ orbitals,
which in turn will give rise to low energy shape resonances. 

\begin{figure}
\newpage
\includegraphics[width=15.5cm]{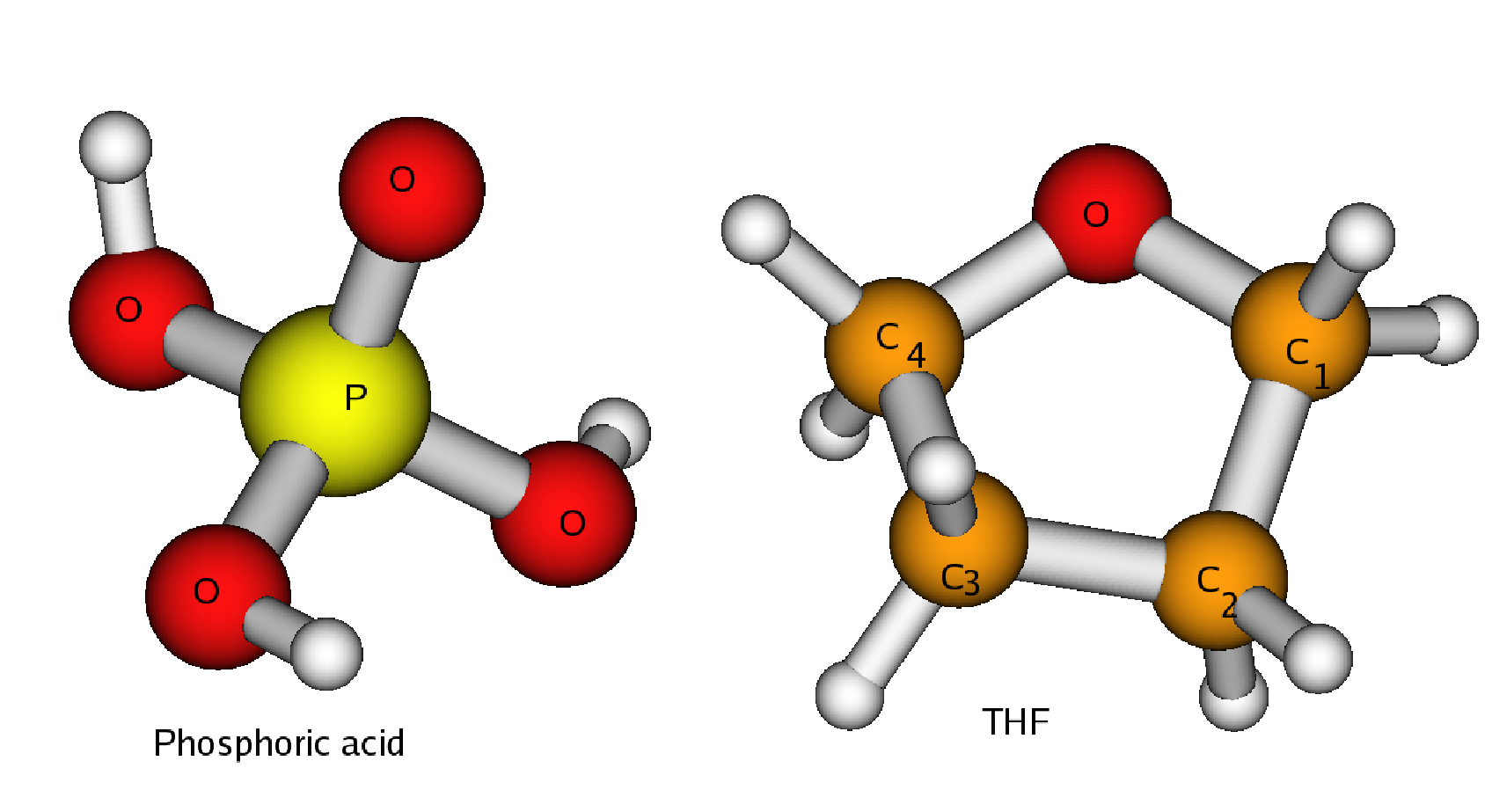}
\newpage
\begin{spacing}{2}
\caption{(Color online) Three dimensional structures of phosphoric acid and
tetrahydrofuran.
The small circles are hydrogen atoms.\label{fig:molecs}
 }
\end{spacing}
\end{figure}

\begin{figure}
\newpage
\includegraphics[width=15.5cm]{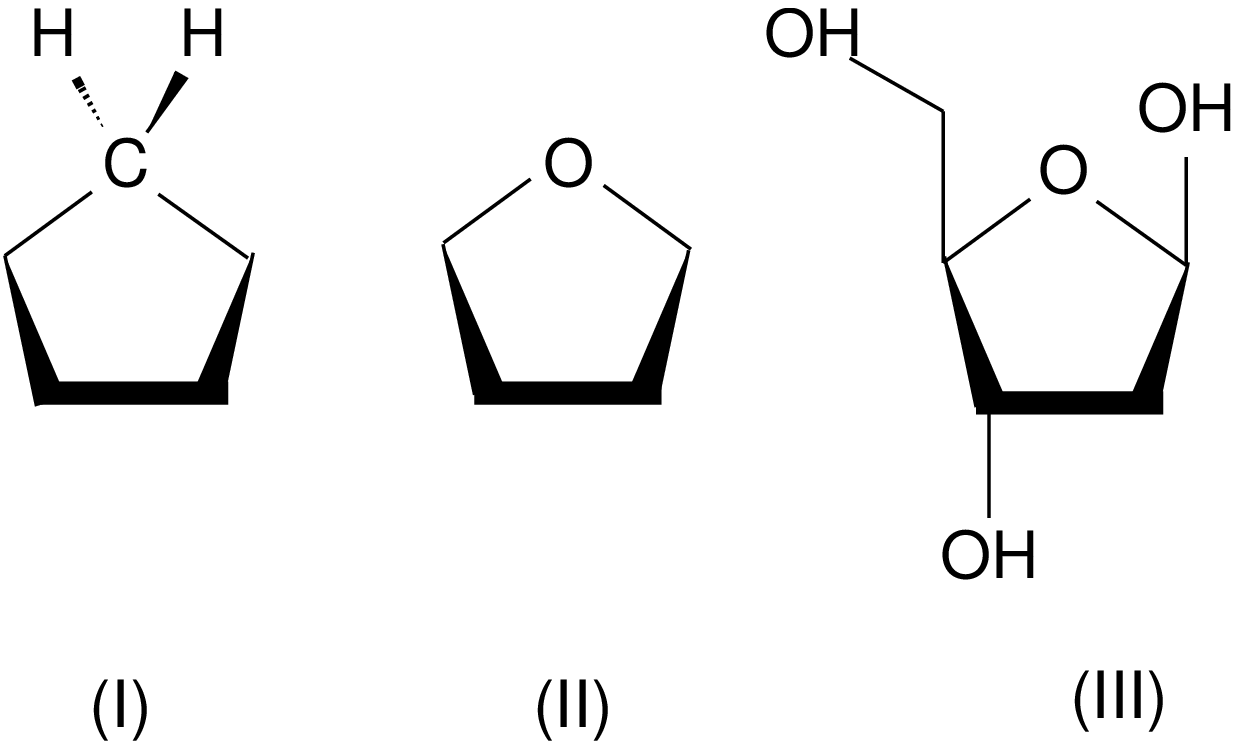}
\newpage
\begin{spacing}{2}
\caption{Schematic structures of cyclopentane (I),  THF (II), and the DNA sugar
deoxyribose (III), that show the similarities between these compounds. The
hydrogen atoms that fill the carbon valences are not shown. The ring
conformation we have used in the calculations is not planar but
puckered, as
indicated in Fig. \ref{fig:molecs} for THF, therefore the C$_{2}$ carbon is
below the plane and C$_{3}$ above, here they are indicated both above for ease
of drawing.\label{fig:molecs_schematic}
}
\end{spacing}
\end{figure}

In a previous article, we
have explored the resonances that arise in electron scattering from each of the
RNA and DNA bases.
\cite{Tonzani:JCP06} In the present study, we analyze the behavior of the
other subunits in the macromolecule, with the aim of investigating the electron
interaction with the moieties that constitute
the sugar-phosphate backbone, and verify whether they could play a role in the
electron capture stage of the process, In any case, it has been shown
\cite{Scheier_ribose:JCP04} that low energy electrons can damage the DNA sugar,
deoxyribose (2-deoxy-D-ribose), and this interaction could be important in the cell environment. 
To model the sugar, we use tetrahydrofuran
(THF, whose formula is C$_{4}$H$_{8}$O), which is similar to deoxyribose, except that the latter
has -OH groups
attached to the C$_1$ and $C_3$  and a CH$_2$OH side chain attached to the C$_4$ (the latter two
substituents are linked to a phosphate group in the DNA molecule, while the former is linked to a base), they are substituted
here by hydrogens. 
We use THF instead of deoxyribose because in DNA the OH groups are
fundamentally modified by the phosphate and base that are attached to it (in
fact the one attached to the C$_{1}$ disappears altogether), so we think that
the solution would be either to consider a whole nucleotide 
or use THF which mimics only the ring structure, which is less affected by the
rest. Another reason for this
choice is that a recent study on dissociative electron attachment (DEA) from
all the DNA components \cite{Burrow:PRL06} has measured the average gas-phase DEA cross
section from the bases
as being very similar to the cross section $per$ $base$ for supercoiled DNA measured in Ref.
\onlinecite{Sanche:rad_research}. This implies the DEA cross section for the sugar
and phosphate groups to be much smaller, 
which is what happens in the case of THF, but not for a hydroxy-substituted
THF, for which the DEA cross section is larger than for the bases.
This suggests that THF could be a better molecule to model the sugar moiety in DNA.
The structures of these compounds are represented schematically in
Figs. \ref{fig:molecs}, \ref{fig:molecs_schematic}. In
practice, the DNA backbone can be thought of approximately as constituted by THF molecules to which
the bases are attached, linked
together by phosphate groups.

For the phosphate there are no previous results available for comparison, but for THF
recently there have been both new experimental
\cite{Brunger:JPB05,Marinkovic:EPJD05} and theoretical \cite{Gorfinkiel:JPB06}
results, with which we will compare our calculations. 
We will also attempt a comparison of the resonant wavefunctions to virtual
orbitals from a Hartree-Fock calculation, in order to gain some more insights
into the capture process and its possible consequences on the anion evolution. 
The conclusions suggested by this analysis lead us to draw a possible link
between our calculations
and  experimental data on dissociative electron attachment .

\section{Theory}
Electron scattering from a polyatomic molecule is intrinsically a  many-body
problem that can, under the so-called static-exchange approximation,\cite{Morr_Coll:PRA78} be reduced
to a one electron problem. The static-exchange approximation amounts to including only the
ground state of the target in the close coupling expansion of the wave
function, and it is essentially the equivalent of the Hartree-Fock
approximation for continuum states. \cite{Froese-Fischer:book} 
A detailed description of our method can be found in Refs.
\onlinecite{Tonzani:JCP05,Tonzani:JCP06,Tonzani:CPC06}, therefore here we will just
sketch the main points of the treatment.

To solve this one electron problem, \cite{Tonzani:JCP05} we use
the R-matrix method which starts by partitioning space into two
zones: a short range zone where all the channels are coupled and the scattering
problem can in principle  be treated in all its many body complexity, and an outer
zone (external to the target electron density) in which the escaping electron only sees the effect of the target molecule as a
multipole expansion of its electrostatic potential.
In its eigenchannel form, the R-matrix method can be formulated as a
variational principle \cite{Greene:rev96} for the normal logarithmic derivative
(-$b$) of the
wavefunction on the reaction zone surface:
% Variational principle for b
\be
\label{log:der}
{b\equiv - \frac{\partial{\log {(r\Psi)}}}{\partial r} = 2 \frac
{\int_V {\Psi^{*}(E-\hat{H} -\hat{L}) \Psi dV}}{\int_V{\Psi^{*}
\delta(r-r_0)\Psi dV}}}
\ee
where $\hat{L}$ is the Bloch operator, needed to make the Hamiltonian $\hat{H}$
Hermitian  and $r_{o}$ is the
boundary between the internal and external regions.
It is possible, after expanding the internal region wavefunction in a
suitable basis set, to recast the solution of Eq. \ref{log:der}
as an eigenvalue problem:
% Eigenvalue problem 
\be
\label{eigenvalue}
\underline{\Gamma}\vec{C}=2({E\underline{O}-\underline{H}-\underline{L}}) \vec{C} =
\underline{\Lambda} \vec{C} b,
\ee
where $\underline{O}$ is a matrix of volume overlap between the basis
functions, 
while $\underline{\Lambda}$ is an overlap of the basis
functions on the surface of the internal zone, \cite{Greene:rev96} essentially
the denominator of Eq. \ref{log:der}, and $\underline{H},\;\underline{L}$ are the
matrix representations of the operators in Eq. \ref{log:der}.
Through basis set partitioning we shift the computational burden to the
solution of a large linear
system.
As a basis set we use finite elements \cite{FEMBEM:notes} in all three
spherical coordinates, in this way we have large but sparse matrices that are amenable
for solution with fast sparse solvers. \cite{Tonzani:CPC06,pardiso}

To simplify further the description of our system we have to deal with the
nonlocality inherent in the potential. To do this we use a local
density approximation for the exchange potential, which reduces it to a
functional only of the local density:
%Maybe can put in general theory
% Local potential
\be
\label{Pot:exch}
V_{ex}(\vec{r}) = -\frac{2}{\pi} k_{F} F(k_{F},E),
\ee
where $k_{F}$ is the local Fermi momentum:
\be
k_{F}(\vec{r})=[3 \pi ^{2} \rho(\vec{r})]^{1/3}
\ee
and $F$ is a functional of the energy and the local
density $\rho(\vec{r})$ (through the local Fermi momentum). The  functional
form we use for $F$ is called the Hara
exchange. \cite{Hara:69}  It has been extensively employed in continuum state
calculations, and it is energy-dependent.  The local exchange approximation, widely used also in
density functional (DFT) calculations, \cite{Parr_Yang:book} has proven itself to give qualitatively correct
results, \cite{Tonzani:JCP05,Morr_Coll:PRA78} while being sufficiently simple
to implement computationally that it permits an exploration of complex
molecular species.

A polarization-correlation potential is added to this.
The long range part of this potential is
a simple multipole expansion, of which we retain only the induced dipole
polarization term: 
\be
\label{polar:potential}
V_{pol} = -\frac{\alpha_{0}}{2 r^{4}}
%+\alpha_{2} P_{2}(\cos{\theta}))
\ee
where $\alpha_{0}$  is
%and $\alpha_{2}$ are 
the totally symmetric 
%and nontotally
component of the polarizability tensor, and it can be calculated $ab$ 
$initio$ using electronic structure codes. Exploratory tests suggest that the
anisotropic polarizability and the electron-quadrupole interaction are much
less important. For example the anisotropic part of the polarizability, when
introduced generates a maximum difference of 0.01 rad in the phase shifts, which
in turn translates in a variation of the cross section values of roughly 1\%.

In the volume where the electronic density of the target is not negligible,
this potential is nonlocal. The polarization-correlation interaction can be approximated again as a
local potential, different forms of which have been suggested in the
literature.  The one
we use is based on DFT (specifically on the LYP potential of
Ref. \onlinecite{Lee_Yang_Parr:PRB88}) and it has yielded
reliable results in the work of Gianturco and coworkers. \cite{Gianturco:c60}
This form makes use of the electron density, its gradient and Laplacian, which have
to be calculated for each target molecule. The short and long range potentials are
matched unambiguously (continuously but with discontinuous derivatives) at the innermost crossing point, whose radius is dependent on the
angles. The matching is unambiguous in the sense that there are two crossing
points between the inner and outer potential for each angle, and we always
choose the innermost, since the other is far in the region where the electron
density of the molecule is very small. Choosing the outermost crossing has proven to give
unphysical results \cite{Gianturco:PRA93} in many cases.
In recent years, better density functionals have been devised that give  the quantities of interest more
directly, without the need for an additional
polarization contribution, and they have been recently used in connection with
time-dependent DFT to calculate scattering observables.
\cite{Burke_Wasserman:JCP05}

Since the molecules considered in this work possess dipole moments, we take dipole interaction
effects into account
in the manner described in Ref. \onlinecite{Tonzani:JCP06}. Recently we discovered
a mistake in our dipole matching code that led to an overestimation of the cross
sections in Fig. 3 of Ref. \onlinecite{Tonzani:JCP06} of about 10\% at very low
energy, decreasing and becoming negligible beyond about 10 eV, but its effects
on the other
results in that work are negligible. The quadrupole fields can potentially be
significant as well, but
they are roughly half of the dipole field at the box boundary; accordingly, we did not
include quadrupole interactions outside the box, since their effect was found
to be small when analyzed using an
R-matrix propagation technique. \cite{Burke_rmatprop:CPC82,Mehta:06} 
The nonspherical part of the polarizability
tensor is quite small as well, roughly one fourth of the spherical part, and
from previous calculations we have seen how such values have only a minimal influence
on the final results; this too has been neglected here.

For molecules with dipole moments the fixed-nuclei scattering cross sections
are formally infinite. This divergence can be readily eliminated by considering rotations of
the molecule, as a Born closure
expansion. \cite{Gianturco:water_JCP98} We have not implemented  this here, as in our
previous work on the DNA bases, because this level of detail is not our present
interest. Moreover,  the dipole moments here are also less than half what they
were for the DNA and RNA
bases, so the correction will be even less important here. Therefore the cross sections
we show should be understood to include only up to a maximum electronic orbital angular momentum
$l_{max}=10$, with all  the higher partial waves omitted.
We stress as well that since our model does not include excited
states of the target molecule, when the electron energy is above any excitation
or the ionization threshold
(the latter is at roughly 9.8 eV for both molecules in this study, but since
our model is approximate we can expect that this will be shifted upward by a
few eV), 
the electron molecule compound will have many more channels to decay into, with
the result that the higher energy resonances will be modified by these new
interactions.

All the target quantities are calculated at the Hartree-Fock level using a
6-31G** basis set, and the target equilibrium geometries have been optimized at the
same level of theory.
The remaining details of the calculations are very
similar to those reported in Ref. \onlinecite{Tonzani:JCP06}, including the
convergence criteria. The dimension of the matrices is
roughly 200000 by 200000, and our  convergence studies show that
increasing the number of sectors by 30\% lowers the position of the resonances
by about 0.1 eV in THF. These calculations are very cumbersome, and for the level of
accuracy we are aiming for here this  convergence criterion seems adequate.  
For details on the treatment of the integrals and of the nuclear Coulomb
singularities we refer the reader to our previous publications, specifically Sec. IIE of Ref.
\onlinecite{Tonzani:JCP05} and Sec. IID of Ref. \onlinecite{Tonzani:JCP06}.

\section{Results: THF}

\begin{figure}
\newpage
\begin{picture}(620,620)(0,0)
\put(0,0){\includegraphics[width=15.0cm]{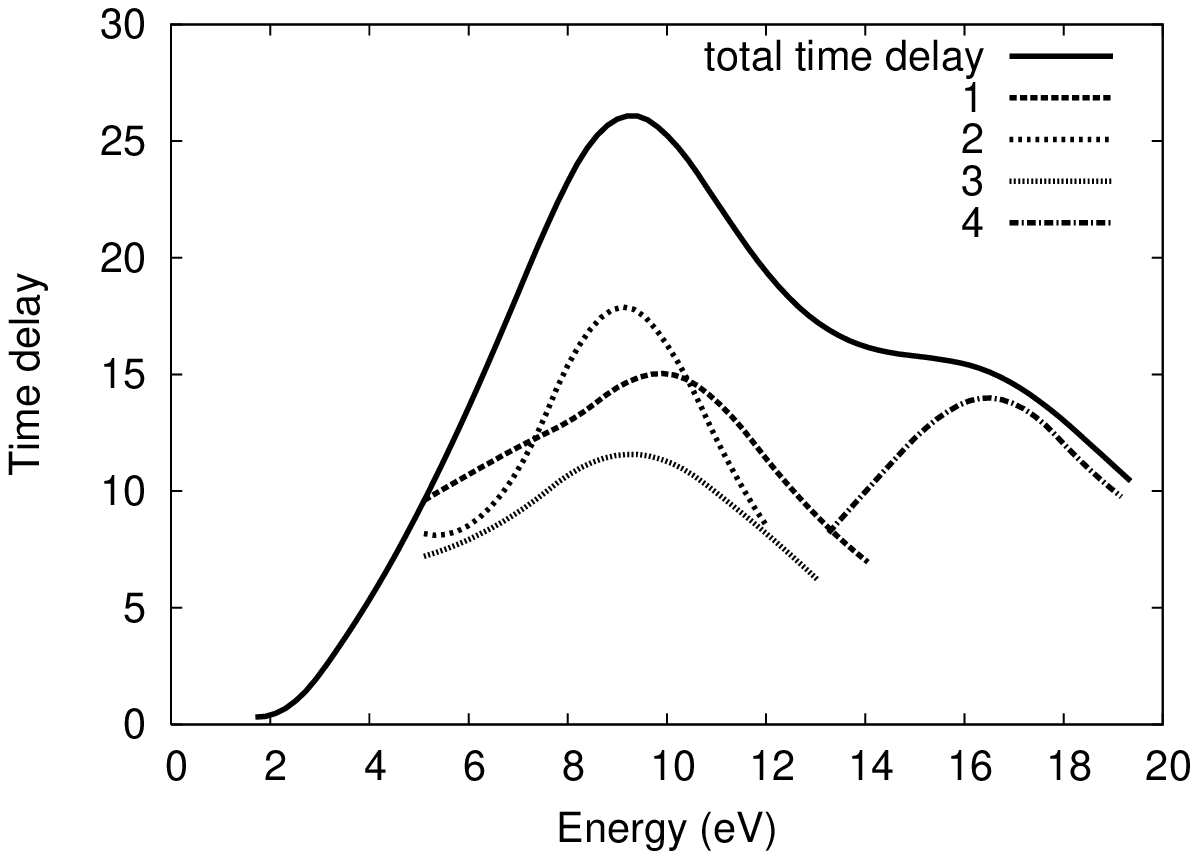}}
\put(0,310){\includegraphics[width=15.0cm]{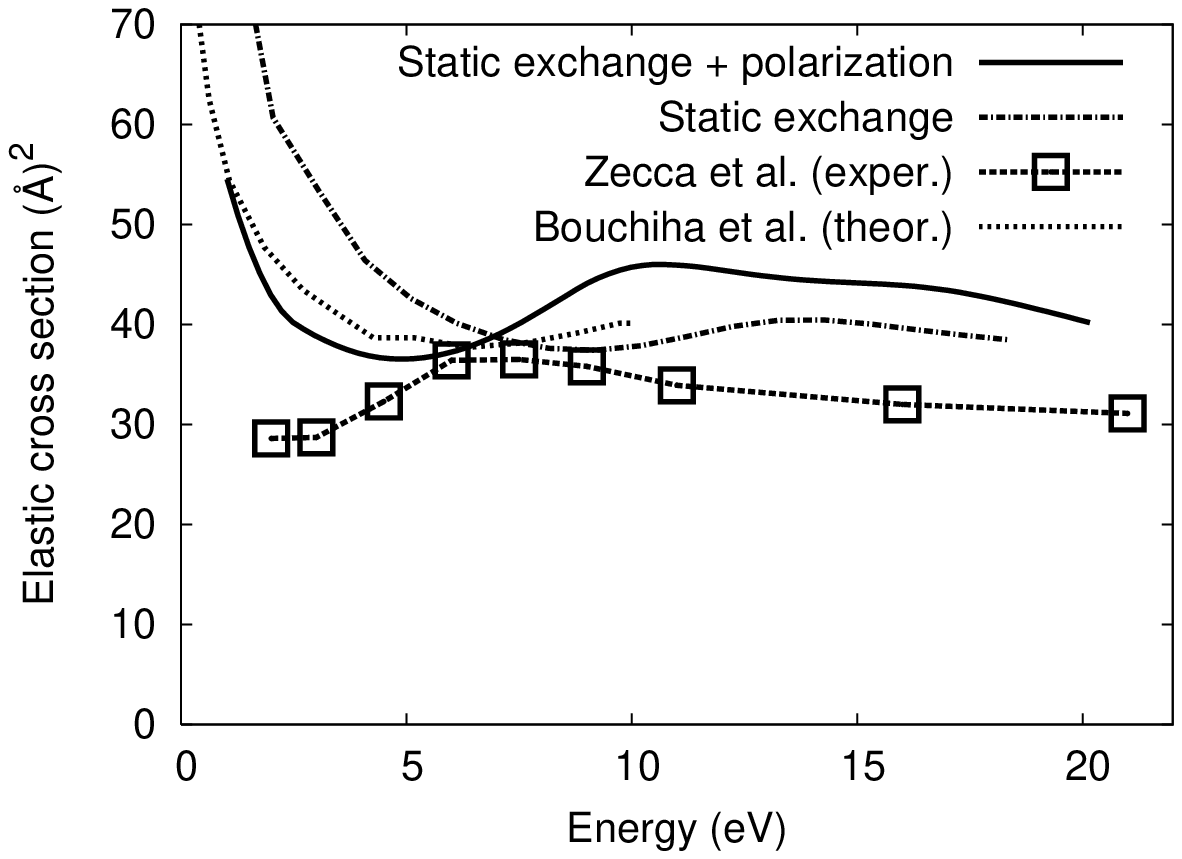}}
 \end{picture}
\newpage
\begin{spacing}{2}
\caption{Partial elastic electron scattering cross sections from THF (solid
line), the dot-dashed line represents the static-exchange results.
Calculations involve partial waves up to l=10 and the
dipole physics outside the R-matrix box is included exactly for those partial
waves.
 Top: cross section comparison with theoretical (dotted) \cite{Gorfinkiel:JPB06} and
experimental (dashed) \cite{Brunger:JPB05} results, the open squares are the actual
experimental data points. Bottom: 
time-delay plot to highlight the presence of resonances. The full curve is the
total time delay, while the numbered curves correspond to the few highest eigenvalues
that exhibit resonant behavior. The total time delay was rescaled to show all
curves on the same graph more easily.
\label{fig:comparison}}

\end{spacing}
\end{figure}

In Fig. \ref{fig:comparison} our results are compared to the low-resolution experimental data
of Zecca $et$ $al.$ \cite{Brunger:JPB05} and the theoretical results of
Bouchiha $et$ $al.$, \cite{Gorfinkiel:JPB06} obtained like ours without
performing a
Born closure  to consider the effect of the
dipole field  on the higher partial waves. We also plot the data in the
form of a time-delay, \cite{Tonzani:JCP06} to make the resonance position and width
more evident.   
The total time-delay is the sum of the eigenvalues of the hermitian matrix
\be
Q=i  S \frac{dS^{\dagger}}{dE},
\label{eq:time_delay}
\ee
where $S$ is the scattering matrix. In Fig. \ref{fig:comparison} we also plot
the resonant channels, particular eigenvalues of the time-delay matrix that
are larger than the others and show a Lorentzian behavior that sets them apart
from the rest. We 
rescaled the total time-delay by dividing it by a factor, to show more easily
everything in the same graph.

The order of magnitude of our cross section is not
different from the results of Ref. \onlinecite{Gorfinkiel:JPB06} calculated
without Born closure. Since our calculations do not include excited states of the target, we will not
be able to detect core-excited resonances (present instead in the calculations
of Ref. \onlinecite{Gorfinkiel:JPB06}). Also comparing with total cross sections
we will be missing rotational and vibrational excitations and the
electronically inelastic channels, therefore elastic cross sections should be
lower than experimental total cross sections, this does not happen in our case,
and it is partly due to the approximate nature of our model.
With respect to experiment, the comparison is difficult
due to the effect of higher partial waves (with $l>10$). 
The Born correction has been
calculated in Ref. \onlinecite{Gorfinkiel:JPB06}, and the cross section becomes
higher than experiment, we would expect a similar effect for our data. Due to the low resolution,
the experimental results below 5 eV are (in the words of the authors) to be
considered only indicative.
The resonant structure that appears around 9.3 eV in
our results looks similar to the analogous experimental feature, also taking into account
that our model usually predicts resonances about 1.5-2 eV higher than their
experimental position, as verified for the DNA bases \cite{Tonzani:JCP06} and
also for molecules like carbon dioxide, benzene and SF$_6$,\cite{Tonzani:unpublished} because of the approximations
adopted.  This is due mostly to the adoption of a local exchange model, since
for example, in a system like CO$_{2}$
\cite{Lane:rev80,Gianturco_Stoecklin:JPhysB96} this approximation causes a
resonance shift of around 2 eV or so for model static-exchange, compared to
all-electron static-exchange calculations, that is not possible to compensate
with the polarization-correlation potential, which indeed shifts the resonance
by a similar amount for this molecule (1.5 eV compared to 2 eV) with respect to correct polarization. Also the model
polarization is approximate, but the model exchange constitutes the largest
error in our calculations. Indeed, if we perform a purely static-exchange
calculation (see Fig. \ref{fig:comparison}) the position of the resonance (maximum of the time-delay) is at
11.9eV, 2.6 eV higher than the result including polarization, and it is 
composed of three overlapping resonances. For what we said
above about the shifts generated by the model potential with respect to exact
exchange, this resonance should be around 10 eV in an all-electron calculation,
therefore probably visible in the calculations of Ref. \onlinecite{Gorfinkiel:JPB06}
that go up to 10 eV in energy. When polarization-correlation is added this
value should fall below 10eV, and thus be clearly visible in the data of
Ref. \onlinecite{Gorfinkiel:JPB06}.

From the fact that our resonances are shifted too high in energy one might have
thought that 
the widths would turn out to be larger than the experimental ones. We have verified in the past
for many systems \cite{Tonzani:JCP06,Tonzani:unpublished}
that our calculated widths are comparable to their correct values.

No resonance was
found in this energy range in the other published theoretical study. \cite{Gorfinkiel:JPB06} 
Through a Lorentzian fit, we predict a
width of about 2.5 eV for this
resonance. At the static-exchange level, the resonance has a larger width, of
about 3.1 eV.  
A higher energy resonance is found at 16.2 eV and it is even broader (about
3.0 eV).
The partial wave contributions are mainly $l=4$ (with a contribution
of 70\%) for the lower energy
resonance and for the higher energy resonance $l=5$ 
(50\%) and $l=3$  (30\%, as can be seen in Tab. \ref{tab:resonances}).
The partial wave decomposition will depend somewhat on what center in chosen as
the expansion. We always perform the decomposition around the center of mass of
the molecule, which seems a reasonable choice.

To understand better the reason for the difference between our results and the ones
in Ref. \onlinecite{Gorfinkiel:JPB06} and to make sure that our
results are consistent with other known results, we have first of all
performed a time-delay analysis \cite{Smith_timedel:PR60,Greene:rev96} of our
scattering data, shown on the bottom of Fig. \ref{fig:comparison}.
This has allowed us to establish the presence of three superimposed resonances,
two of which peak at 9 eV (with one sensibly more intense than the other) while the third
peaks around 10eV, and its intensity falls between the other two. 
Due to the large width of these resonances, the weakest of the channels in Fig.
\ref{fig:comparison} has a time-delay
eigenvalue which is not much larger than the nonresonant eigenvalues.

An existing calculation for cyclopropane \cite{Gianturco:JPB02} (C$_{3}$H$_{6}$)
 shows that this cycloalkane has a cross section with a
similar resonant structure at 6 eV. It is superimposed on a broader resonance
at higher
energy and other nonresonant contributions that give a very wide
plateau, although in this case the first feature is not a composite resonance.
The symmetry of the cyclopropane resonance in Ref. \onlinecite{Gianturco:JPB02} is even with respect to the plane of the three
carbon atoms, and the same occurs  in THF, in the sense that although this compound
is not planar, our resonance
structures do not change sign above and below the ring bonds. It is plausible then that due to
its larger frame and density of virtual orbitals, THF is able to support
more 
quasi-degenerate resonant states at a similar energy. 

\begin{figure}
\newpage
\centerline{\includegraphics[width=15cm]{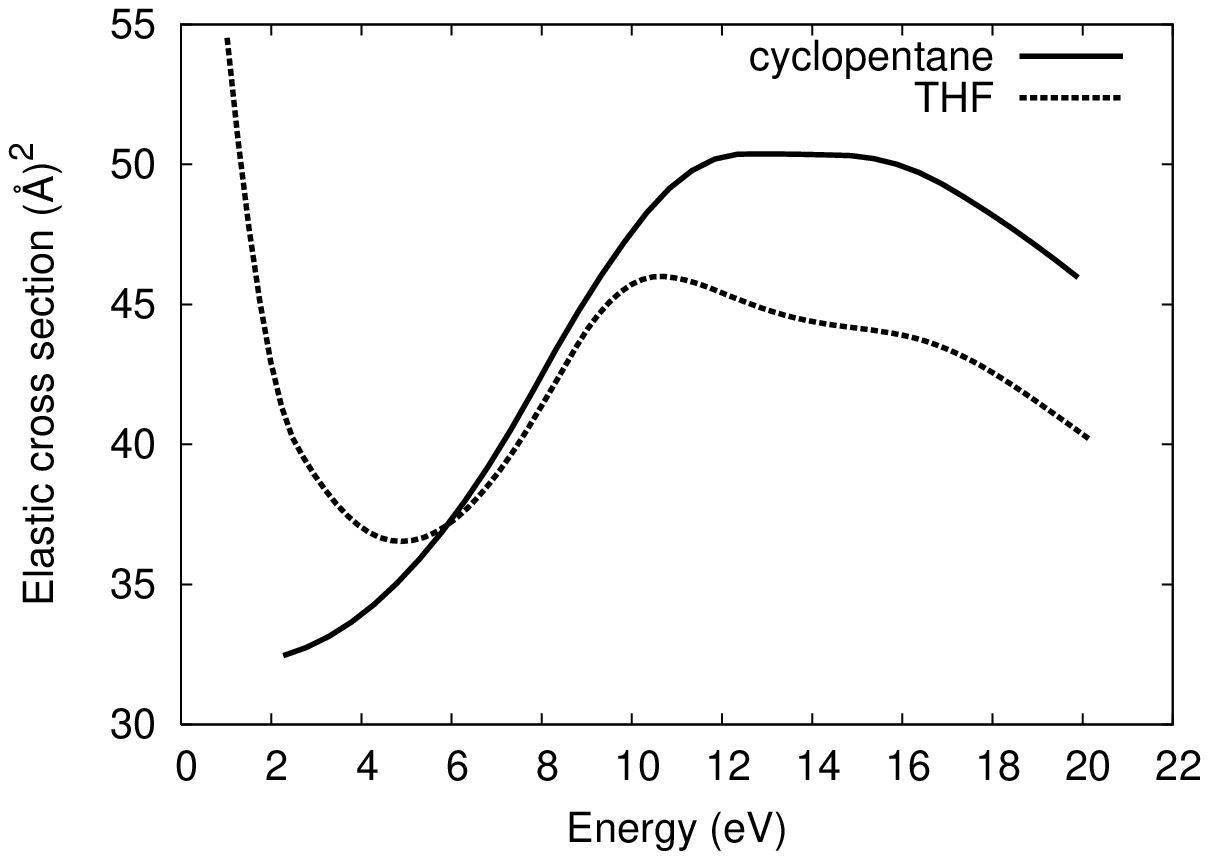}}
\newpage
\begin{spacing}{2}
\caption{Unconverged fixed nuclei elastic cross sections for electron scattering
by THF and
cyclopentane. The two molecular structures are similar (THF has an oxygen atom instead of a
-CH$_2$- group) and so are their electron scattering  cross sections.
Cyclopentane has essentially zero dipole moment, therefore the low energy part
of its cross section does not rise the way it does in THF.
\label{fig:THF_cyclopentane}}

\end{spacing}
\end{figure}

Since there are no other existing experimental data on THF, 
we have performed
calculations for  cyclopentane (C$_{5}$H$_{10}$), which is very similar to THF except that here 
the oxygen atom is substituted with a -CH$_2$- group. 
For this molecule, experimental data \cite{Sueoka:ACP00} shows a broad shape
resonance at 8eV. The similarity of the two molecules suggests that many
features of the cross section will be analogous.
The results are
compared to THF in Fig. \ref{fig:THF_cyclopentane}. The cross section closely
resembles that of
THF, with
a triple resonance superimposed on a broad higher
energy resonance centered around 16 eV, which produces a wide plateau. The electron
scattering cross sections for alkanes and cycloalkanes of different sizes
have very similar behaviors,
\cite{Sueoka:ACP00,Sueoka:PRA05} namely the same resonant structures and a magnitude
dependent on the number of carbon atoms. Our evidence thus suggests that THF, being very
similar to a cycloalkane, shares these common features, as suggested also in
Ref. \onlinecite{Brunger:JPB05}. Also the cross section
for cyclopentane is, in our calculations, larger than in THF, a trend that
seems to be confirmed by the experimental
data.\cite{Brunger:JPB05,Sueoka:PRA05} The cyclopentane cross section peak is roughly 50
\AA$^2$ which is very close to the experimental value \cite{Sueoka:ACP00} for
the total cross section. Recent and as yet unpublished
theoretical calculations, \cite{Trevisan_Orel_THF:06} performed on a THF
molecule distorted to achieve a planar geometry for the furanose ring,
show results similar to
ours, with a double shape resonance in a similar energy range.

\begin{figure}
% OLD FIGURE
%\begin{picture}(500,600)(0,0)
%\put(0,240){ 
%\includegraphics[width=8.5cm]{figures/THF_5delay.0.05.trans.eps}
%}
%\put(240,0){
%\includegraphics[width=7.5cm,angle=90]{figures/THF_orbital31.ps}
%}
%\put(0,460){
%\includegraphics[width=8.5cm]{figures/THF_3delay.0.3.transp.eps}
%}
%\put(240,240){
%\includegraphics[width=6.5cm,angle=90]{figures/THF_orbital27.epsi}
%}
%\put(240,420){
%\includegraphics[width=7.5cm,angle=90]{figures/THF_orbital23.ps}
%}

\newpage
\begin{picture}(400,500)(0,0)
\put(0,0){ 
\includegraphics[width=8.5cm,height=9.0cm]{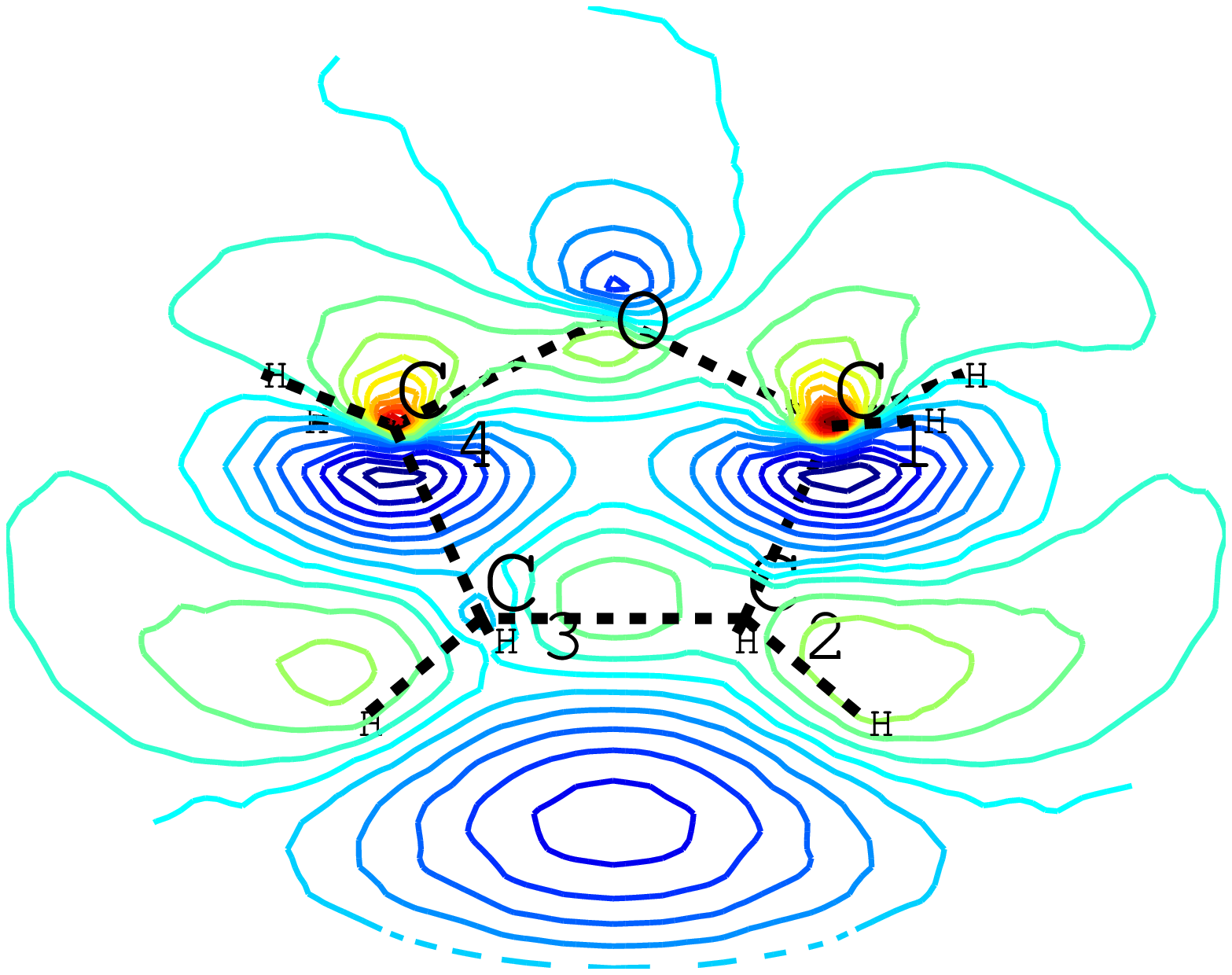}
}
\put(250,10){
\includegraphics[width=8.0cm]{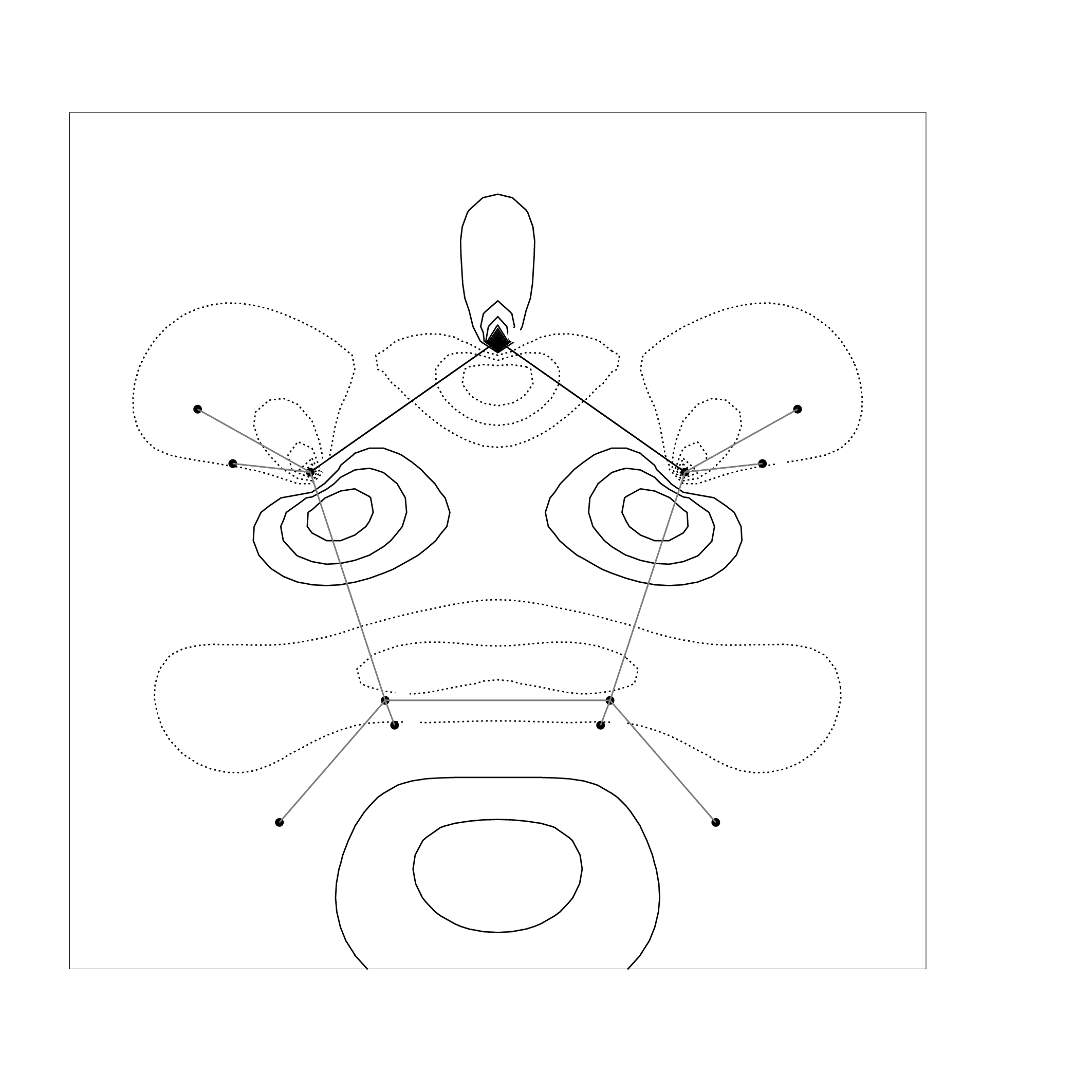}
}
\put(-20,240){
\includegraphics[width=9.5cm,height=9.0cm]{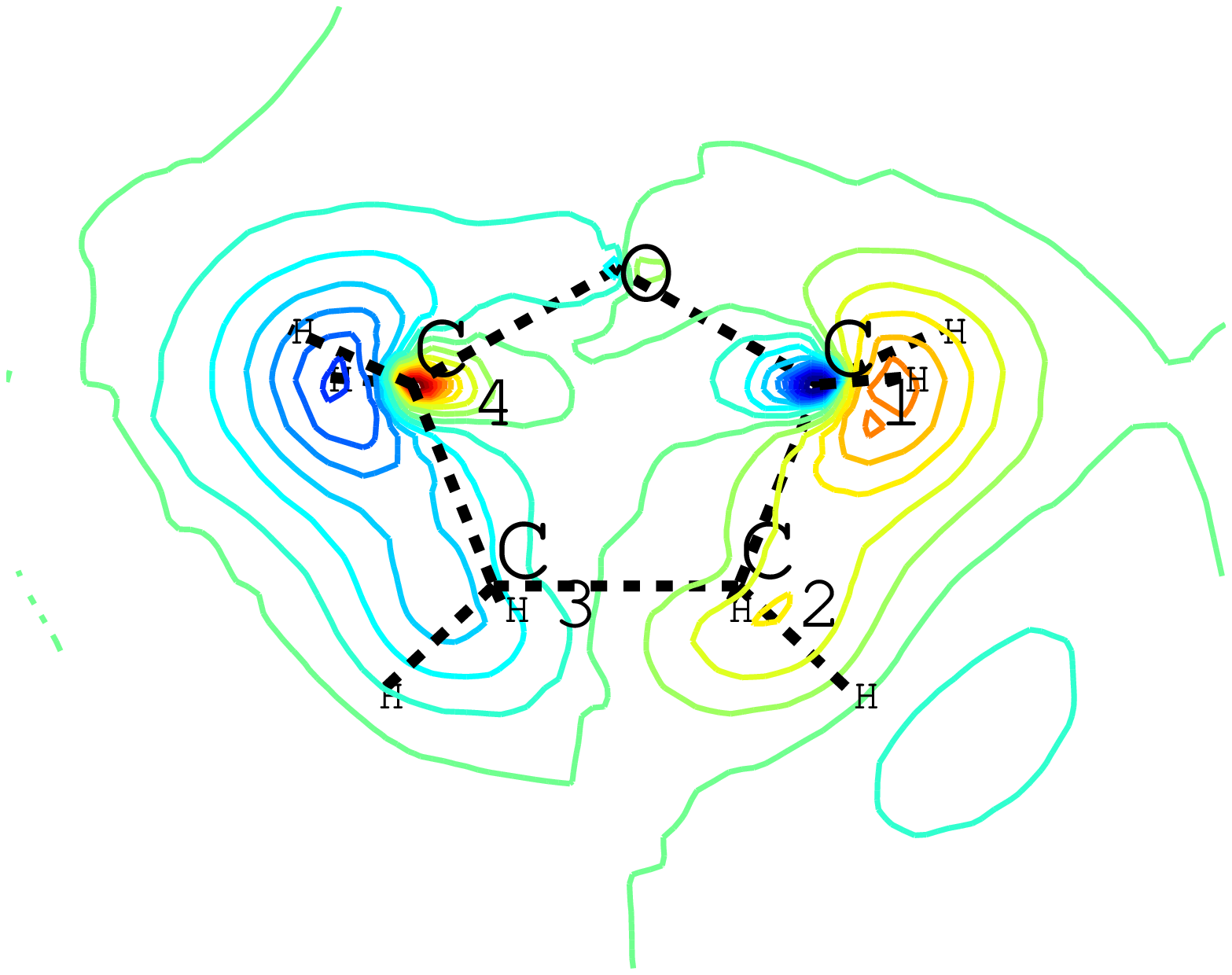}
}
\put(250,260){
\includegraphics[width=8.0cm]{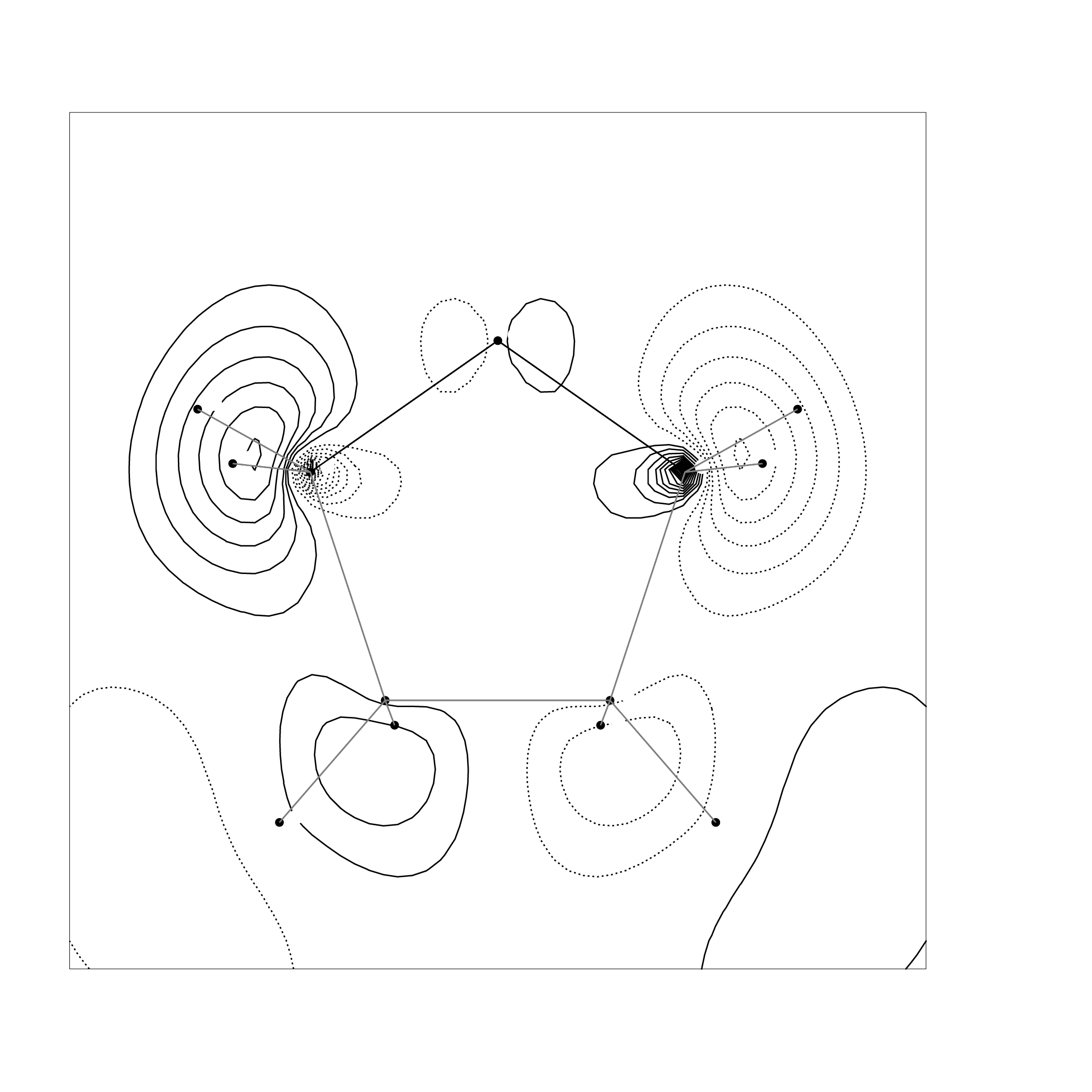}
}
\end{picture}
\newpage
\begin{spacing}{2}
\caption{(Color online) Left: Time delay eigenfunctions  for THF on resonance
at 9eV (top, where the widely-spaced dotted curve (labeled "2") in Fig. \ref{fig:comparison} is dominant)
and 10.5eV (bottom, where instead the dashed curve (labeled "1") in Fig. \ref{fig:comparison}
dominates). A slice of the three dimensional eigenfunctions is shown,  
on the plane that contains OC$_{1}$C$_{4}$, while C$_{3}$ is above
the plane and C$_{2}$ below. The red contours identify positive areas and the
blue negative areas of the real part of the wavefunction. 
Right: virtual orbitals for
energies around 7 (top) and 11 eV (bottom) from HF at the 6-31G** level. The
full lines identify positive areas and the broken lines negative areas.
The correspondence between the two top plots and the two bottom plots is very
pronounced, allowing to identify the two main contributions to the resonance.
Although the molecule is not planar this projection is much easier to read
than the three dimensional structures. \label{fig:resonances}
}

\end{spacing}
\end{figure}

As in Ref. \onlinecite{Tonzani:JCP06} we attempt to correlate the spatial shape of the
resonant wavefunction with the Hartree-Fock virtual orbitals obtained with a
small (6-31G**) basis set. In this case the analysis is complicated by the fact
that the molecules are not planar, so a projection on a
two-dimensional surface would not pass through all the nuclei and, since the resonance is wide, 
different contributions mix and overlap. However, we have established that the
resonant wavefunction at its peak resembles closely a virtual orbital with energy of 7
eV (orbital 23).  At higher energy (around 10.5 eV, when the second resonant contribution
becomes dominant), the resonance spatial structure  appears similar to an orbital
of 11 eV of energy
(orbital 31). We show this in Fig. \ref{fig:resonances}. Since there is
a great degree of resonance overlap, the  relationship of the resonant
wavefunctions to the virtual orbitals we show becomes even clearer when observing the other
virtual orbitals, which are actually very different from our wavefunctions.

Recently, DEA cross sections have been
measured \cite{Burrow:PRL06} for all the DNA subunits. These cross sections display two
prominent resonances at 6 and 8 eV, of width around 1 eV each for THF, which
could correlate with the calculated shape resonance, although clearly much
complicated vibrational dynamics happens between electron capture and
dissociation that we do not take into account here, and in the results of Ref.
\onlinecite{Gorfinkiel:JPB06} there are electronically inelastic resonances in this energy region
which could lead to similar outcomes in DEA.

Several electron-energy-loss (EEL) studies of THF
\cite{Sanche_Lepage:JCP98,Bremner:FD91} and cyclopentane
\cite{Allan_cyclopentane:JCP96} have been
performed, and some attempts have been made at assigning the resonances, also in
relation to the position of the excited states from absorption spectroscopy. In
Ref. \onlinecite{Bremner:FD91} a gas-phase EEL peak for THF at 6.6 eV is assigned as a
core-excited resonance with a parent ion of Rydberg character, centered on the
oxygen atom, whereas a similar experimental peak in cyclopentane in Ref.
\onlinecite{Allan_cyclopentane:JCP96} is assigned as a shape resonance, on the
basis of a molecular orbital argument, with the electron  wavefunction
distributed around the
whole ring. Our calculations seem to agree with this second assignment, for two
reasons: first, the
resonance is very similar in character in THF and cyclopentane (see Fig.
\ref{fig:THF_cyclopentane}), and therefore it seems implausible that it derives
from excitations on the oxygen; second, the resonant wavefunctions
in  Fig. \ref{fig:resonances} are found to be delocalized over the entire ring
as had been speculated in Ref. \onlinecite{Allan_cyclopentane:JCP96} for cyclopentane. At higher
energies core-excited contributions are most likely predominant, as was suggested
in Ref. \onlinecite{Sanche_Lepage:JCP98}.

It is difficult to attempt a prediction of what the evolution of
the resonances found in elastic scattering could be, when the nuclear dynamics is included. At present
this is computationally prohibitive, but a first analysis can be attempted by
inspecting our calculated results.
From the
spatial structures of the resonant wavefunctions  it seems
likely that the resonances could initiate a break-up of the ring, because of the presence of
many nodal surfaces cutting through the ring bonds. For example, the
experimental DEA cross
sections for deoxyribose in Ref.
\onlinecite{Scheier_ribose:JCP04} exhibit a peak at around 6 eV for production
of the anion
C$_{3}$H$_{5}$O$_{3}^{-}$, which could be generated by a resonant structure
that shows a node cutting through the ring from C$_{3}$ to C$_{1}$, similar to
the top left plot in Fig. \ref{fig:resonances}, given that deoxyribose has side
groups attached to C$_{4}$ and C$_{3}$ that could lead to production of the
aforementioned anion.

%%%%%%%%%%%%%%%%%%%%%%%%%%%%%%%%%%%
% Resonance table
\begin{table}
\newpage
\begin{tabular}{|l|c|c|c|} \hline
Molecule & Energy (eV) & Width (eV) & Partial wave \\ \hline
\emph{THF}   &9.3  &2.5  &4 ($70\%$)  \\ 
  &16.2  &3.0  &5 ($50 \%$)  \\ \hline

\emph{H$_{3}$PO$_{4}$}   &9.1  &2.0  &4 ($42\%$)  \\ 
  &15.9  &1.5  &4 ($41 \%$)  \\ \hline

\end {tabular}
\newpage
\begin{spacing}{2}
\caption{Energies, widths and dominant partial waves of the resonances
discussed in the text.\label{tab:resonances} }
\end{spacing}
\end{table}
%%%%%%%%%%%%%%%%%%%%%%%%%%%%%%%%%%%

%%%%%%%%%%%%H3PO4%%%%%%%%%%%%%%%%%%%%

\section{Results: H$_3$PO$_4$}
\begin{figure}
\newpage
\begin{picture}(620,620)(0,0)
\put(0,0){
\includegraphics[width=15.5cm,height=10cm]{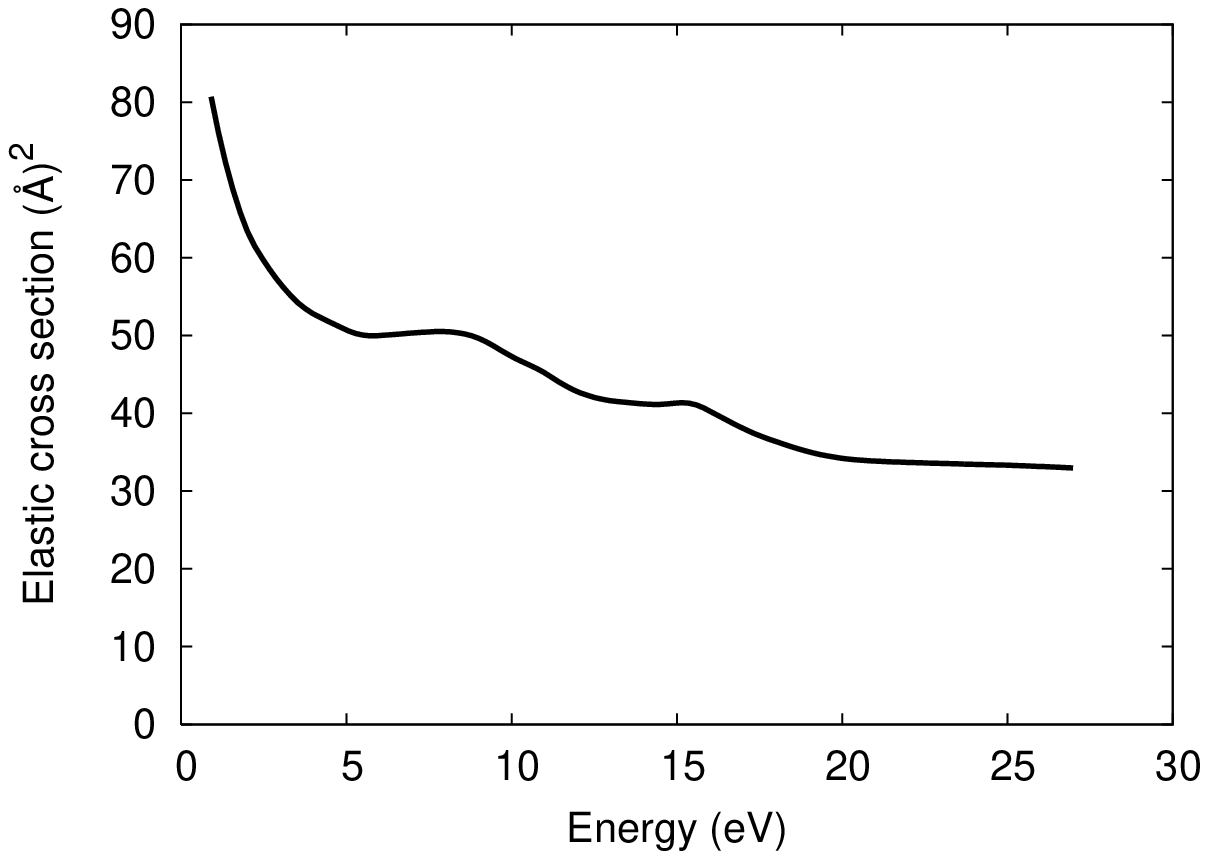}}
\put(0,310){\includegraphics[width=15.5cm,height=10cm]{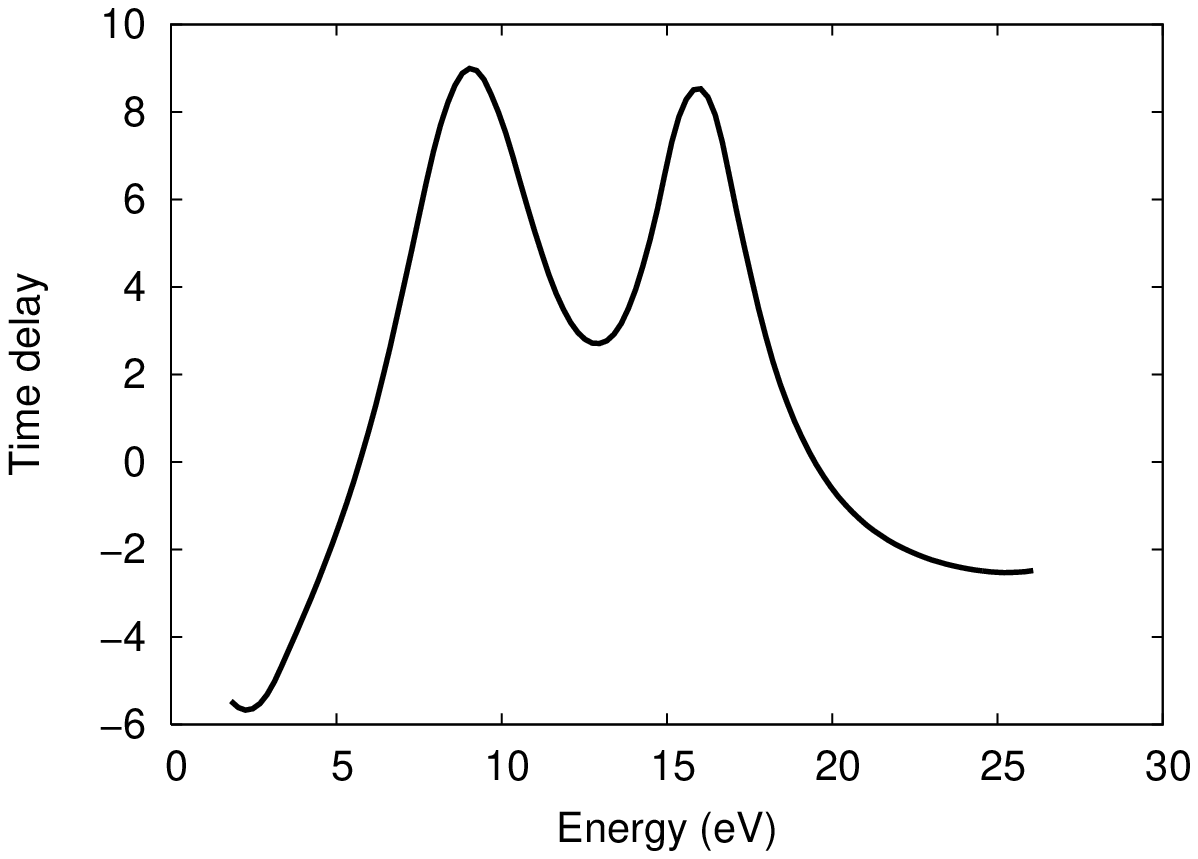}}
\end{picture}
\newpage
\begin{spacing}{2}
\caption{H$_3$PO$_4$ partial elastic cross section (top) and time-delay analysis (bottom). Calculations again involve partial waves up to l=10 and the
dipole physics outside the R-matrix box is included exactly for those partial
waves. Two broad resonances are present, at 8.9 and 15.7 eV, the  cross section is smaller than
in the DNA bases and comparable to THF. \label{fig:H3PO4_cross}
}
\end{spacing}
\end{figure}

In the case of phosphoric acid there are no available theoretical or
experimental data to compare with, and consequently our analysis of the results will
be somewhat more limited. The 
cross sections show two prominent resonances, one at 9.1 eV and the other
at 15.8 eV, with respective widths 2.0 eV and 1.5 eV. 
The main partial wave contributions are $l=4$ and $l=3$ at 40\%
and 30\% respectively for both resonances, as stated in Tab.
\ref{tab:resonances}. At low energies there is a sharp rise in cross section
due to the dipole moment. Cross section and total time delay are illustrated in
Fig. \ref{fig:H3PO4_cross}.
It is difficult to say why the second resonance is narrower than the first,
since this is somewhat counterintuitive. We will limit ourselves to pointing out
that in a complicated polyatomic molecule,
shape resonance widths depend not only on the total energy, but also the size
of the probability density near the escape regions in the multidimensional
potential. It is possible, therefore, that a higher-energy resonance could be
narrower than one at lower energy.
Resonances broader at lower energy than the higher energy ones can be seen,
for example, in C$_{60}$ fullerene.
\cite{Gianturco:c60,McKoy:c60}

In Ref.
\onlinecite{Burrow:PRL06} the DEA cross section has been measured for
trimethylphosphate (PO$_4$(CH$_3$)$_3$), and it displays a very wide and prominent resonant structure at 7.5 eV, 
to which our 9.1 eV resonance could correlate. No branching ratio has been measured
for this compound. The structure of phosphoric acid presents three
identical -OH groups and therefore the resonances will show  structures with
a similar probability for the scattering electron to end up in each of these
groups, making for complicated spatial profiles, which in the end are not very
illuminating.
For these reasons we will not attempt here an analysis of the spatial structure of the resonances.

As in THF, the shape resonances are quite broad, so the anions will be
relatively short lived, and probably they will autoionize back (possibly with
vibrational excitation) before having
the possibility to lead to a breakup or energy transfer to another DNA subunit.
However, the presence of water, a structural component in biological DNA, can  act to stabilize the anions, and
probably will significantly influence the lifetimes of these resonances in the
cell nucleus, since the backbone is in closer contact with the water and the
surrounding environment, compared to the bases that lie inside the double
helix and are presumably somewhat less affected by the presence of the solvent. 
 
\section{Conclusions}
We have calculated electron scattering cross sections from molecules that closely resemble the
DNA sugar-phosphate
backbone components, namely tetrahydrofuran and phosphoric acid, as a first step to model
the DNA radiation damage caused by low energy electrons. Since our model is approximate, resonance shifts of
around 2 eV higher than experiment are usually observed. In THF we find a
very broad resonance, which agrees (within the limits of our model) with recent experimental data, but not with
one recent theoretical calculation. Also for phosphoric acid we find shape
resonances. In both cases the resonances are quite broad and high in energy and
are not expected to play a major role, compared to those of the nitrogenous bases, in electron capture leading to damage in
DNA. For THF we have also attempted to correlate the resonant structures with
recent DEA experiments and to electron scattering experiments from other
related molecules.

\section*{Acknowledgments}
This work was supported by the Department of Energy, Office of Science, and
NERSC with an allocation of supercomputing resources. We thank P. Burrow, 
J. Gorfinkiel, L. Sanche and L. Caron for stimulating discussions, the group of J. Toomre for
computational time on their machines and N. Mehta for the use of his R-matrix propagator
computer code.

%\section{References}
\bibliography{paper_FEA}
\printfigures
\end{document}